\documentclass[ reprint,
  amsmath,amssymb, aps,pra]{revtex4-1}
\usepackage{tikz}
\usepackage{amscd}
\usepackage{amsthm}
\usepackage{graphicx}
\usepackage{mathdots}
\usepackage{epstopdf}
\usepackage{enumerate}
\usepackage{changepage}

\usepackage[
colorlinks,
linkcolor = blue,
citecolor = blue,
urlcolor = blue]{hyperref}
\def \qed {\hfill \vrule height7pt width 7pt depth 0pt}

\newtheorem{theorem}{Theorem}
\newtheorem{corollary}{Corollary}
\newtheorem{lemma}{Lemma}

\newtheorem{proposition}{Proposition}

\usepackage{dcolumn}
\usepackage{bm}

\begin{document}

\preprint{APS/123-QED}

\title{Twist-teleportation-based local discrimination of maximally entangled states}
\author{Mao-Sheng Li$^{1,2}$}
\author{Shao-Ming Fei$^{3,4}$}\email{feishm@cnu.edu.cn}
\author{Zong-Xing Xiong $^{5}$}
\author{Yan-Ling Wang$^{6}$} \email{wangylmath@yahoo.com}

\affiliation{
 $^1$Department of Physics,
 Southern University of Science and Technology, Shenzhen 518055, China}
\affiliation{
 $^2$ Department of Physics, University of Science and Technology of China, Hefei 230026, China}

\affiliation{ $^3$School of Mathematical Sciences, Capital Normal University,
Beijing 100048, China}

\affiliation{ $^4$Max-Planck-Institute for Mathematics in the Sciences, 04103
Leipzig, Germany}

\affiliation{ $^5$Department of Mathematics,
 South China University of Technology, Guangzhou
510640,  China}

\affiliation{
 $^6$School of Computer Science and Network Security, Dongguan University of Technology, Dongguan 523808, China}

\begin{abstract}
In this work, we study the local distinguishability of maximally entangled states (MESs). In particular, we are concerned with whether any fixed number of MESs can be locally distinguishable for sufficiently large dimensions.
Fan and Tian \emph{et al.} have already obtained two satisfactory results for the generalized Bell states (GBSs) and the qudit lattice states when applied to prime or prime power dimensions.
We construct a general twist-teleportation scheme for any orthonormal basis with MESs that is inspired by the method used in [Phys. Rev. A \textbf{70}, 022304 (2004)]. Using this teleportation scheme, we obtain a sufficient and necessary condition for one-way distinguishable sets of MESs, which include the GBSs and the qudit lattice states as special cases.
Moreover, we present a generalized version of the results in [Phys. Rev. A \textbf{92}, 042320 (2015)] for the arbitrary dimensional case.
\begin{description}
\item[PACS numbers] 03.67.Hk, 03.65.Ud, 03.65.Aa
\item[Key words] teleportation, local discrimination, maximally entangled states
\end{description}
 \end{abstract}                            
\maketitle

\section{Introduction}
In quantum information processing, it is often found that the subsystems of a composite system are spatially separated. Therefore, quantum manipulation of the system can only be conducted using local operations and classical communication (LOCC).
The local distinguishability of quantum states plays important roles in the exploration of LOCC capabilities \cite{Bennett99,Walgate02}. If we suppose that Alice and Bob share a bipartite quantum state chosen from a set of previously known orthogonal states, then their task is to identify the given state using only LOCC.
The set of states are said to be locally distinguishable or distinguishable by LOCC if LOCC protocols exist to exactly identify the states of the given set; otherwise, the set is locally indistinguishable or indistinguishable by LOCC or nonlocal. The study of local distinguishability of quantum states has direct applications in data hiding \cite{DiVincenzo02} and quantum secret sharing \cite{Rahaman15}.
	
Since Bennett \emph{et al.} discovered a locally indistinguishable $\mathbb{C}^3\otimes\mathbb{C}^3$ pure product basis in \cite{Bennett99}, considerable research has focused on identifying the sets of orthogonal product states or the maximally entangled states (MESs) that are locally indistinguishable \cite{Ben99b,Hor03,DiVincenzo03,Ran04,Feng09,Yang13,Zhang14P,Zhang15P,Wang15,Zhang16P,Xu16m,Zhang17P,Halder18,Walgate00,Ghosh01,Ghosh04,
 Duan12,Cosentino13,Bandyopadhyay13,CosentinoR14,Li15,Yu15,Yu14,Bandyopadhyay15,Bandyopadhyay11,Nathanson2013,Zhang14,Zhang15,Tian15_2,Duan12,Cosentino13,CosentinoR14,Li15,Yu15,Yang18,Wang19,Xiong19}.
One reason for performing these constructions is that by constructing locally indistinguishable orthogonal quantum states, it is possible to better understand the boundary between LOCC operations and global operations.

However, its applications in data hiding and quantum secret sharing require participants to reveal the encoding results using LOCC.
In other words, quantum states used to hide secrets should be identifiable under LOCC. Therefore, in addition to constructing sets of quantum states that cannot be locally distinguished, it is also crucial to study the sufficient condition to ensure that a set of quantum states can be locally distinguished.

Numerous results have been obtained for the local distinguishability of MESs. In 2004, Fan \cite{Fan04} noticed that their local distinguishability is not changed under local unitary operations. Using a series of complex Hadamard matrices acting on the generalized Bell states (GBSs), it has already been successfully proven that any $l$ GBSs of prime $d$-dimensional  can be locally distinguished provided $l(l-1)\leq 2d$. In 2005, Nathanson showed that any three orthogonal MESs can be distinguished by LOCC in $\mathbb{C}^3\otimes\mathbb{C}^3$, which is also thought to be true for higher-dimensional systems \cite{Nathanson05}.
Fan's result was extended by Tian \textit{et al.} to the prime power dimensional mutually commuting qudit lattice states \cite{Tian15}. The classification of the local distinguishability of four GBSs in $\mathbb{C}^4\otimes\mathbb{C}^4$ has been analyzed in \cite{Singal15,Tian16}. Wang \emph{ et al.} showed that any three orthogonal GBSs can be locally distinguished for any dimension $d\geq4$ \cite{Wang17}. It is of interest to investigate whether Fan's and Tian \emph{ et al.}'s results  can be extended to arbitrary dimensional quantum systems.
	
This article is organized as follows. In Sec. \ref{sec}, we  discuss the relationship between MESs and unitary matrices. We also review some important MESs such as the GBSs and the qudit lattice states. In Sec. \ref{thi}, for a given maximally entangled basis, we present a direct proof of the teleportation scheme over unknown channels. In Sec. \ref{fou}, we present a     sufficient  and necessary condition for a set of special MESs to be distinguished under one-way LOCC. In fact, we show that one-way LOCC discrimination of special MESs is equivalent to the distinguishability of the corresponding unitary matrices. We also study the properties of this equivalence problem. In Sec. \ref{fiv}, we apply the main results obtained in Sec. \ref{fou} to general qudit lattice states. We obtain a general version of the results given by Fan and Tian \emph{et al.} for any dimensionality. Furthermore, for a given $l$, we consider whether any $l$ qudit lattice states can be distinguished by LOCC for a large enough dimension $d$. Finally, we conclude and propose some interesting problems in Sec. \ref{last}.

\section{ Unitary matrices and MESs }\label{sec}

Consider a $d\times d$ bipartite quantum system $\mathcal{H}_A\otimes \mathcal{H}_B$. Let $\{|0\rangle, |1\rangle, \cdots, |d-1\rangle \}$ be the computational basis of the subsystem.
The standard MES can be expressed as ${|\Psi_0\rangle=\frac{1}{\sqrt{d}} \sum_{i=0}^{d-1}|ii\rangle}$.
Any MES can be uniquely written in the form $|\Psi_U\rangle=(U\otimes I)|\Psi_0\rangle$ for some unitary matrix $U$.
There is a one-one correspondence between MESs in $\mathcal{H}_A\otimes \mathcal{H}_B$ and unitary matrices in $\text{U}(d)$.
We call $U$ the corresponding unitary matrix of $|\Psi_U\rangle$. Moreover, the following relationship applies:
$$\langle \Psi_V|\Psi_U\rangle= \text{Tr}(V^\dagger U)=\langle V, U\rangle.$$
That is, the correspondence is inner-product preserving. Below are two sets of important MESs: the GBSs and the qudit lattice states.

\begin{enumerate}[(i)]
  \item \textbf{GBSs.}
  Let $d$ be an integer with $d\geq 2$ and $\omega_d=e^{\frac{2\pi \sqrt{-1} }{d}}$ be a primitive $d$th root of unity.
We define the bit-flip and phase-flip operators to be:
{\small$$ X_d=\displaystyle\sum_{i=0}^{d-1}|i+1 \text{ mod } d\rangle\langle i|, \text{ and } Z_d=\displaystyle\sum_{i=0}^{d-1}\omega_d^i|i\rangle\langle i|.$$ }
The following $d^2$ orthogonal MESs are called the GBSs:
{\begin{equation}\label{eq1}
\{|\Psi_{m,n}\rangle=(X_d^mZ_d^n\otimes I)|\Psi_0\rangle\big|  0\leq m,n\leq d-1\}.
\end{equation}
}
Noting that $Z_dX_d=w_dX_dZ_d$, it follows that:
{\begin{equation}\label{commutative1}
(X_d^mZ_d^n)(X_d^{m'}Z_d^{n'})= \widetilde{w}(X_d^{m'}Z_d^{n'})(X_d^mZ_d^n),
\end{equation}}
where $\widetilde{w}= w_d^{m'n-mn'}$.
This implies that the defining matrices of GBSs are commutative up to some phases.
  \item  \textbf{{Qudit lattice states.}} Firstly, we consider a simple case $d=p^r$ where $p$ is  a prime. Let $\mathbb{Z}/p\mathbb{Z}=\{0, 1,...,p-1\}$ be the additive group with $p$ elements. For any $r$ dimensional vectors $\textbf{s}=(s_1,s_2,...,s_r)$ and
        $\textbf{t}=(t_1,t_2,...,t_r)$ belonging to $(\mathbb{Z}/p\mathbb{Z})^r$, we define unitary matrices as follows:
       $$X_p^{\textbf{s}} Z_p^\textbf{t}:=(\otimes_{i=1}^rX_p^{s_i})(\otimes_{i=1}^rZ_p^{t_i}).$$
      The qudit lattice states of dimension $p^r$ are defined to be of the form $X_p^{\textbf{s}} Z_p^\textbf{t}\otimes I|\Psi_0\rangle.$
      More generally, let $d=\prod_{j=1}^fp_j^{r_j}$ be the prime decomposition of $d$. Set $\textbf{s}=(\textbf{s}^{(1)},\textbf{s}^{(2)},...,\textbf{s}^{(f)})$ and $\textbf{t}=(\textbf{t}^{(1)},\textbf{t}^{(2)},...,\textbf{t}^{(f)}),$ with $\textbf{s}^{(j)},\textbf{t}^{(j)}\in {(\mathbb{Z}/p_j\mathbb{Z})}^{r_j}.$
We define a lattice unitary matrix to be:
        $$\mathcal{X}^{\textbf{s}} \mathcal{Z}^\textbf{t}:=\otimes_{j=1}^f X_{p_j}^{\textbf{s}^{(j)}} Z_{p_j}^{\textbf{t}^{(j)}}.$$
Further, the qudit lattice states of dimension $\prod_{j=1}^fp_j^{r_j}$ are defined to be of the form $\mathcal{X}^{\textbf{s}} \mathcal{Z}^\textbf{t}\otimes I|\Psi_0\rangle.$ It is easy to check that a similar commutative relation to (\ref{commutative1}) holds:
        {\begin{equation}\label{commutativela}
         (\mathcal{X}^\textbf{s}\mathcal{Z}^\textbf{t})(\mathcal{X}^{\textbf{s}'}\mathcal{Z}^{\textbf{t}'})= w(\textbf{s},\textbf{t},\textbf{s}',\textbf{t}')(\mathcal{X}^{\textbf{s}'}\mathcal{Z}^{\textbf{t}'})
         (\mathcal{X}^\textbf{s}\mathcal{Z}^\textbf{t}),
        \end{equation}
        }
        where $|w(\textbf{s},\textbf{t},\textbf{s}',\textbf{t}')|=1.$
\end{enumerate}

If a set of orthogonal unitary basis $\{U_i\}_{i=1}^{d^2}$ of $\text{M}_d(\mathbb{C}) $ satisfies the following relations:
$$U_iU_j^T=w(i,j)U_j^TU_i, \text{ with } |w(i,j)|=1,$$
then this basis is  called to be  twist commutative. Note that $X_d^T=X_d^{d-1},Z_d^T=Z_d$. Combining these equalities with Eqs. (\ref{commutative1}) and (\ref{commutativela}),
it is easy to ascertain that both the GBS and the qudit lattice basis are twist commutative.

\vskip 6pt

\noindent \textbf{Remark:} If $\mathcal{B}_j=\{U_i^{(j)}\}_{i=0}^{d_j^2-1}$ is a mutually orthogonal unitary basis of $\text{\emph{M}}_{d_j}(\mathbb{C})$ that is twist commutative for $j=1,2,...,f$,
with $d=\prod_{j=1}^fd_j$, then the following set
$$\mathcal{B}=\{U^{(1)}\otimes U^{(2)}\otimes\cdots \otimes U^{(f)} \ \big | \ U^{(j)}\in \mathcal{B}_j, j=1,2,...,f\}$$
is also an orthogonal unitary basis of $ \text{\emph{M}}_{d}(\mathbb{C})$ that is twist commutative.

In this paper, we focus primarily on MESs $\{|\Psi_{U_i}\rangle\}$ that correspond to a twist-commutative unitary basis $\{U_i\}$.
Because there is a one-to-one correspondence of an MES with its defining unitary matrix, we identify a set of MESs with the set of corresponding defining unitary matrices,
{\begin{align}\label{state_study}
 \mathcal{L}:= \{|\Psi_{U_{n_i}}\rangle \ \big |\ 1\leq i\leq l\}= \{U_{n_i} \ \big |\ 1\leq i\leq l\}.
\end{align}
}

\section{Twist quantum teleportation scheme}\label{thi}
Let $|\Psi_0\rangle=\frac{1}{\sqrt{2}}(|00\rangle+|11\rangle),|\Psi_i\rangle=\sigma_i \otimes I|\Psi_0\rangle,$ where $i\in\{x,y,z\}$ and
$$
\sigma_x=\left[
           \begin{array}{cc}
             0& 1 \\
             1 &0 \\
           \end{array}
         \right],
\sigma_y=\left[
           \begin{array}{cc}
             0 & 1 \\
             -1 &0 \\
           \end{array}
         \right],
 \sigma_z=\left[
           \begin{array}{cc}
             1& 0 \\
             0 &-1 \\
           \end{array}
         \right].
         $$
The quantum teleportation \cite{nil} of a qubit state $|\psi\rangle=\alpha|0\rangle+\beta|1\rangle$
is based on the following equation:
$$\begin{array}{rl}
  |\psi\rangle\otimes|\Psi_0\rangle = &\frac{1}{2}(|\Psi_0\rangle\otimes|\psi\rangle+|\Psi_x\rangle\otimes\sigma_x^\dagger |\psi\rangle \\
  &+
|\Psi_y\rangle\otimes\sigma_y^\dagger|\psi\rangle+|\Psi_z\rangle\otimes\sigma_z^\dagger|\psi\rangle).
\end{array}
$$

The above teleportation scheme also works for the $d$-dimensional case \cite{Werner01}.
Suppose $\{|\Psi_i\rangle \ \big | i=0,1,...,d^2-1\}$ is an orthogonal maximally entangled basis of a bipartite system $\mathcal{H}_A\otimes \mathcal{H}_B$ with $\text{dim}_\mathbb{C}\mathcal{H}_A=\text{dim}_\mathbb{C}\mathcal{H}_B=d.$ Without loss of generality, we can assume that $|\Psi_0\rangle=\frac{1}{\sqrt{d}}\sum_{i=0}^{d-1}|ii\rangle$ under the computational basis $\{|i\rangle\}_{i=0}^{d-1}.$ Then there exists an uniquely unitary matrix $U_i$ corresponding
to $|\Psi_i\rangle$,
$$|\Psi_i\rangle=U_i\otimes I |\Psi_0\rangle=I\otimes U_i^T|\Psi_0\rangle.$$

\begin{lemma}\label{tele}
For any pure states $|\psi\rangle_{{\text{\tiny  C}}}=\sum_{i=0}^{d-1}\alpha_i|i\rangle_{{\text{\tiny C}}},$ we have
$$
|\psi\rangle_{{\text{\tiny  C}}}|\Psi_r\rangle_{{\text{\tiny  AB}}}=\frac{1}{d}\sum_{i=0}^{d^2-1}|\Psi_i\rangle_{{\text{\tiny  CA}}}\otimes U_r^TU_i^\dagger|\psi\rangle_{{\text{\tiny  B}}},
$$
where the sub-indices $A,B$, and $C$ denote the qudits $A,B$, and $C$, respectively.
\end{lemma}

\noindent\emph{Proof:} Since $\{|\Psi_i\rangle_{{\text{\tiny  AB}}}\}_{i=0}^{d^2-1}$ is an orthogonal normalized basis of $\mathcal{H}_A\otimes \mathcal{H}_B$,
$
\{|\Psi_i\rangle_{{\text{\tiny  CA}}}|j\rangle_{{\text{\tiny  B}}}\big | 0\leq i\leq d^2-1, 0\leq j\leq d-1\}
$
is an orthogonal normalized basis of $\mathcal{H}_C\otimes\mathcal{H}_A\otimes \mathcal{H}_B$,
\begin{align}\label{tele1}
 |\psi\rangle_{{\text{\tiny  C}}}|\Psi_r\rangle_{{\text{\tiny  AB}}}  = \displaystyle\sum_{i=0}^{d^2-1}\sum_{j=0}^{d-1}  (_{{\text{\tiny  CA}}}\langle\Psi_i | _{{\text{\tiny  B}}} \langle j||\psi\rangle_{{\text{\tiny  C}}}|\Psi_r\rangle_{{\text{\tiny  AB}}})|\Psi_i\rangle_{{\text{\tiny  CA}}}|j\rangle_{{\text{\tiny  B}}}.
\end{align}
The coefficients on the right-hand side of (\ref{tele1}) can be written as:
{ \small\begin{align}\label{tele2}
\begin{array}{cl}
     &       _{{\text{\tiny  CA}}}\langle\Psi_i | \ _{{\text{\tiny  B}}} \langle j|\  |\psi\rangle_{{\text{\tiny  C}}}|\Psi_r\rangle_{{\text{\tiny  AB}}}\\[2mm]
       = &\displaystyle\sum_{k=0}^{d-1} \alpha_k( _{{\text{\tiny  CA}}}\langle\Psi_0 | I \otimes {U_i^T}^\dagger)\  _{{\text{\tiny  B}}} \langle j|\ |k\rangle_{{\text{\tiny  C}}}|\Psi_r\rangle_{{\text{\tiny  AB}}}    \\[4mm]
      =&{\small \displaystyle\sum_{k=0}^{d-1}\sum_{m=0}^{d-1}\sum_{n=0}^{d-1}  \frac{\alpha_k}{d} {}_{{\text{\tiny  CA}}}\langle mm | \ _{{\text{\tiny  B}}}\langle j|(I_{{\text{\tiny  C}}}\otimes {U_i^T}^\dagger U_r \otimes I_{{\text{\tiny  B}}})  |k\rangle_{{\text{\tiny  C}}}|nn\rangle_{{\text{\tiny  AB}}} }\\[4mm]
      =& \displaystyle\sum_{k=0}^{d-1}\sum_{m=0}^{d-1}\sum_{n=0}^{d-1}  \frac{\alpha_k}{d}\delta_{mk} \delta_{jn}\langle m| {U_i^T}^\dagger U_r |n\rangle \\[4mm]
     = &\displaystyle\sum_{k=0}^{d-1}  \frac{\alpha_k}{d}\langle k| {U_i^T}^\dagger U_r |j\rangle.
  \end{array}
\end{align}
}
Clearly,
\begin{align}\label{tele3}
\sum_{j=0}^{d-1}(\sum_{k=0}^{d-1}\alpha_k \langle k| {U_i^T}^\dagger {U_r^T}^T |j\rangle)|j\rangle=U_r^TU_i^\dagger|\psi\rangle.
\end{align}
By combining equations (\ref{tele1}), (\ref{tele2}), and (\ref{tele3}) one proves the Lemma.\qed
\vskip 5pt

Lemma \ref{tele} shows that, if Alice and Bob share the MESs $|\Psi_r\rangle_{AB}$ and Alice wants to teleport the state $|\psi\rangle_C$ to Bob, she only needs to make a projective measurement under the basis $\{|\Psi_i\rangle\}_{i=0}^{d^2-1}$ and to tell Bob the measurement outcome.
However, if Alice and Bob do not know exactly which MES they share, Bob could not recover the state $|\psi\rangle$ perfectly.
However, he knows that his state must be one of the $\{ U_r^T U_i|\psi\rangle\}_{r=0}^{d^2-1}.$
We call such a teleportation scheme twist teleportation (See Fig. \textcolor[rgb]{0.00,0.07,1.00}{1}).
	\begin{figure}[h]
		\includegraphics[width=0.49\textwidth,height=0.37\textwidth]{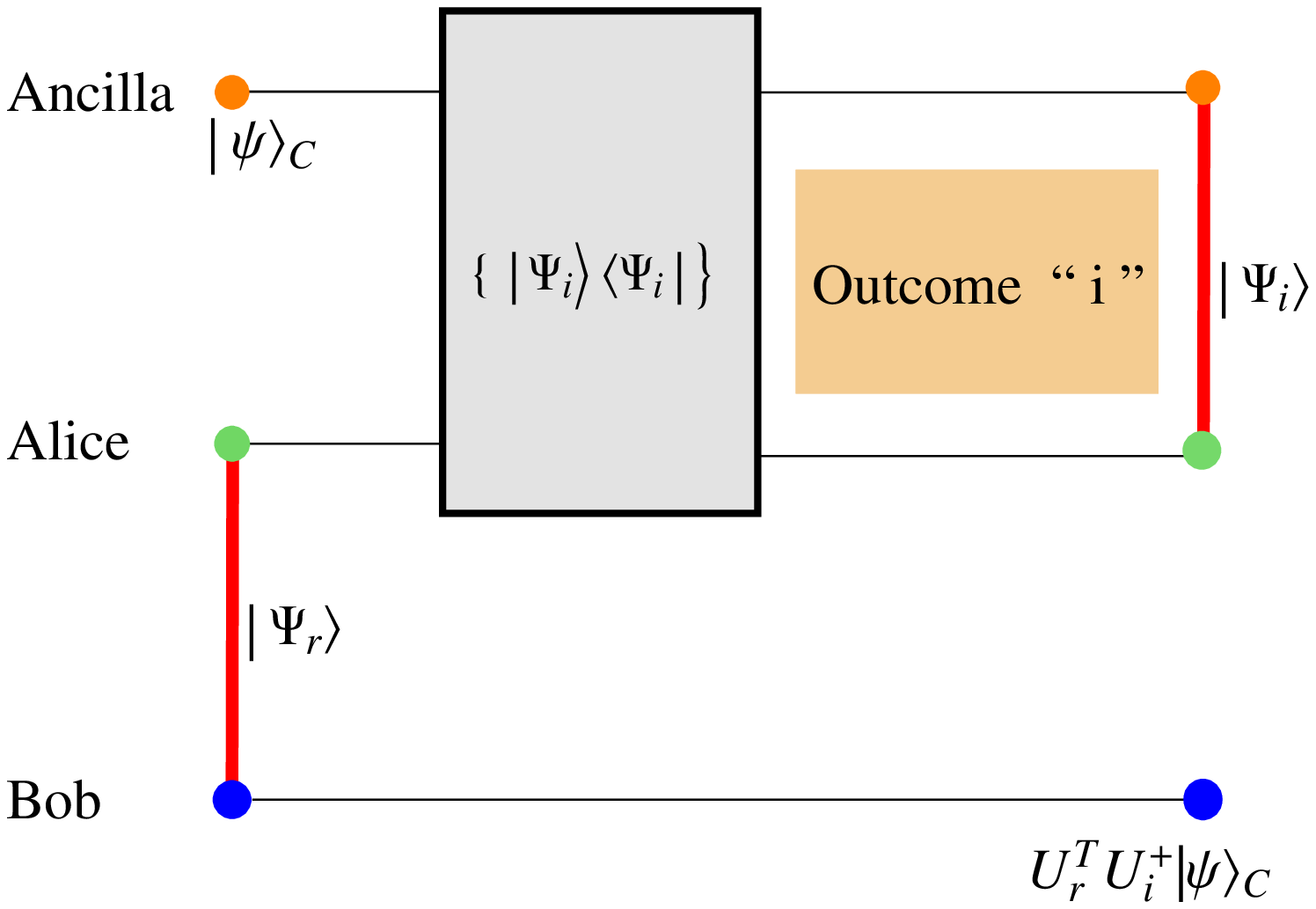}
\begin{adjustwidth}{4 mm}{4 mm}
\noindent{\textbf{Fig. 1:}{\small\emph{ Sketch of twist teleportation  protocol given in Lemma \ref{tele}.    }}}
\end{adjustwidth}	
	\end{figure}

\section{Local distinguishability of twist commutative MESs}\label{fou}

Given a set of unitary operators $\{U_i\}_{i=1}^l\in \text{U}(d)$, we say that they are distinguishable if there exists a unit vector $|\alpha\rangle\in \mathbb{C}^d$ such that $\{ U_i|\alpha\rangle\}_{i=1}^l$ are pairwise orthogonal.
By definition, two nonorthogonal unitary matrices might be distinguishable. For example,
the following two $3\times 3$ matrices
$$
U_1=\left[\begin{array}{ccc}
  1 & 0 & 0 \\
  0 & 1 & 0 \\
  0 & 0 & 1
\end{array}\right],\ \ \
U_2=\left[\begin{array}{ccc}
  1 & 0 & 0 \\
  0 & 0 & 1 \\
  0 & 1 & 0
\end{array}\right]
$$
are distinguishable as it is possible to choose $|\alpha\rangle=(0, 1, 0)^T$.

It should be noted that $\{U_i\}_{i=1}^l\subseteq \text{U}(d)$ are distinguishable if and only if $\{U_i^T\}_{i=1}^l\subseteq \text{U}(d)$. In fact, $\{ U_i|\alpha\rangle\}_{i=1}^l$ are pairwise orthogonal if and only if $\{ U_i^T|\overline{\alpha}\rangle\}_{i=1}^l$ are pairwise orthogonal, where $|\overline{\alpha}\rangle$ is the complex conjugation of $|{\alpha}\rangle$.
The following is a general version of one of the main results (the Theorem 1) in \cite{Ghosh04}.

\begin{theorem}\label{main1}
Let $ \{|\Psi_i\rangle\}_{i=0}^{d^2-1}$ be an orthonormal maximally entangled basis of $\mathcal{H}_A\otimes\mathcal{H}_B$ whose corresponding unitary matrices $\mathcal{S}=\{U_{i}\}_{i=0}^{d^2-1}$ are twist commutative. Let $\mathcal{L} $ be a subset of $\mathcal{S}$. Then, the states corresponding to $\mathcal{L}$ can be distinguished by one-way LOCC from $A\rightarrow B$ if and only if the unitary matrices of the set $\mathcal{L}$ are distinguishable.
\end{theorem}
\noindent\emph{Proof:} The proof of necessity has been investigated by Nathanson \cite{Nathanson2013}. With regard to sufficiency, suppose $|\alpha\rangle$ is a unity vector such that $\{U^T|\alpha\rangle\big|\ U\in \mathcal{L}\}$ are pairwise orthogonal. First, Alice prepares the state $|\alpha\rangle$ in her ancilla system $C$. Then, Alice and Bob use the state $|\Psi_U\rangle$, which needs to be identified as the resource state to teleport Alice's state $|\alpha\rangle$ according to the twist-teleportation scheme shown in Fig. \textcolor[rgb]{0.00,0.07,1.00}{1}. Alice first performs projective measurement under the basis $\{ |\Psi_i\rangle\}_{i=0}^{d^2-1}$. If the outcome is ``$i$", Lemma \ref{tele} tells us that Bob's states are just $U^T U_i^\dagger |\alpha\rangle$. By the commutative assumption, it follows that $U^T U_i^\dagger |\alpha\rangle\propto U_i^\dagger U^T|\alpha\rangle$. Hence, for $U,V\in  \mathcal{L} \text{ and } U\neq V$, we have
$$ |\langle \alpha| U_i(V^T)^\dagger U^T U_i^\dagger |\alpha\rangle|= |\langle \alpha|  (V^T)^\dagger U^T   |\alpha\rangle|=0.$$
This implies that Bob can distinguish these states by a projective measurement according to the orthonormal set $\{U^T U_i^\dagger |\alpha\rangle\}_{U\in \mathcal{L}} $,
see Fig. \textcolor[rgb]{0.00,0.00,1.00}{2}.
\qed

\begin{figure}[h]
\includegraphics[width=0.49\textwidth,height=0.37\textwidth]{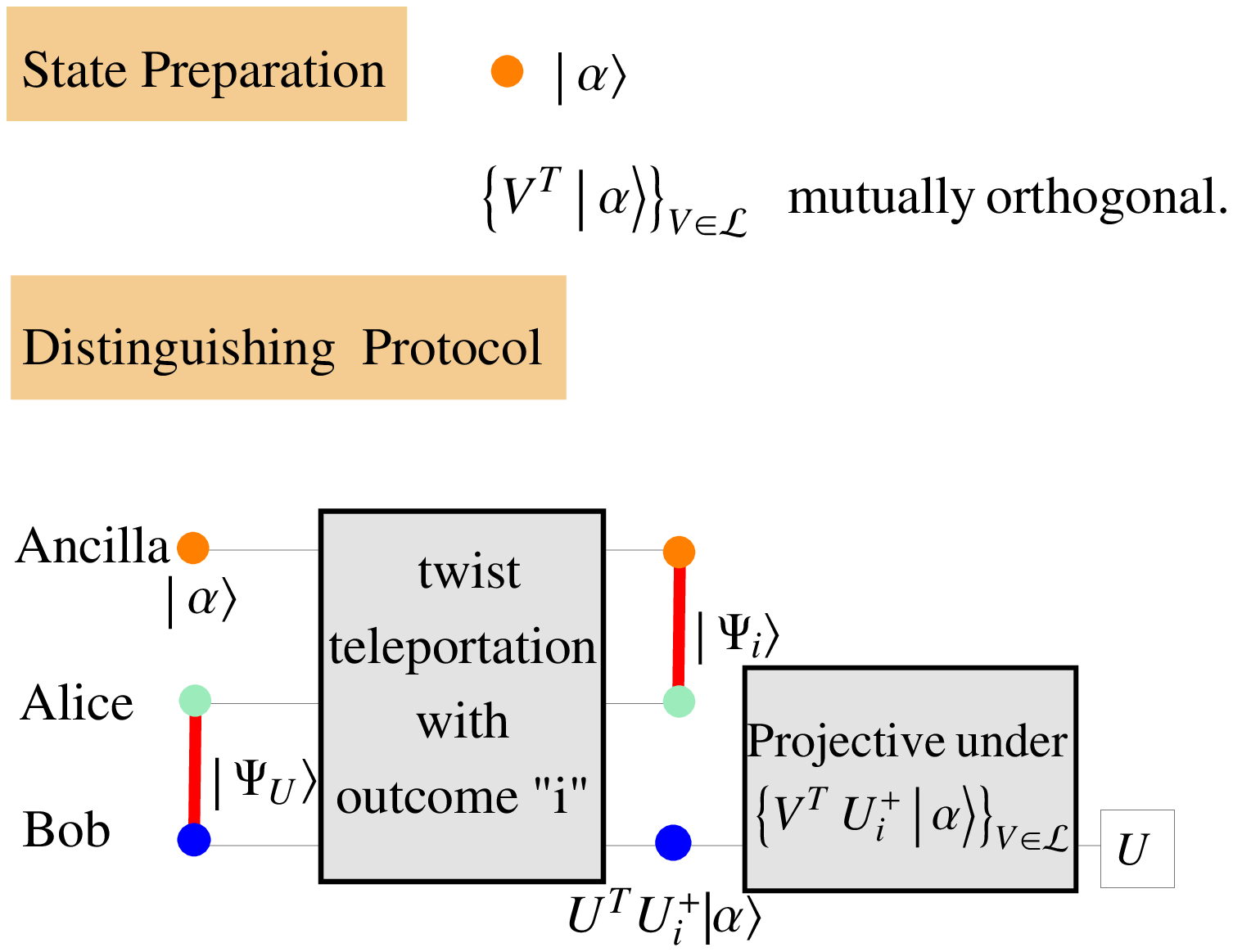}
\begin{adjustwidth}{4mm}{4mm}
\noindent{\textbf{Fig. 2:}{\small\emph{ A sketch of the teleportation protocol based on Theorem \ref{main1}.
There are two stages: first, Alice prepares a state $|\alpha\rangle$ and manipulates the twist-teleportation scheme; then, Bob identifies the state by projective measurement.}}}
\end{adjustwidth}	
	\end{figure}

It might be found that studying whether a set of unitary matrices are distinguishable is more efficient than determining the one-way LOCC distinguishable problem of the corresponding states.
In the following we present some useful properties of the distinguishability of unitary matrices.

\begin{proposition}\label{unitary_tele} Let $d_1$ and $d_2$ be two integers such that $2\leq d_1\leq d_2$. For any orthogonal unitary basis  $\{U_i\}_{i=1}^{d_1^2}$ of $\text{\emph{M}}_{d_1}(\mathbb{C})$ and $V\in \text{\emph{U}}(d_2)$, the set of unitary matrices $ \{U_i\otimes V\}_{i=1}^{d_1^2}\subseteq \text{\emph{U}}(d_1d_2)$   are distinguishable.
\end{proposition}

\noindent\emph{Proof:} Let $\{|i\rangle_A\}_{i=1}^{d_1}$ and $\{|i\rangle_B\}_{i=1}^{d_2}$ be the computational orthonormal bases of systems $A$ and $B$, respectively.
Set $|\alpha\rangle=\sum_{i=1}^{d_1}|i\rangle_A|i\rangle_B$. For $1\leq k,l\leq d_1^2,$ we have
$$\begin{array}{rl}
     &\langle\alpha|(U_k^\dagger\otimes V^\dagger)(U_l\otimes V)|\alpha\rangle\\ [2mm]
        = &  \sum_{i=1}^{d_1}\sum_{j=1}^{d_1} {}_A\langle i|{}_B\langle i| U_k^\dagger U_l\otimes I |j\rangle_A|j\rangle_B \\[2mm]
        = & \sum_{i=1}^{d_1}\sum_{j=1}^{d_1} {}_A\langle i|U_k^\dagger U_l|j\rangle_A \delta_{ij}\\[2mm]
       = & \sum_{i=1}^{d_1}{}_A\langle i|U_k^\dagger U_l|i\rangle_A \\[2mm]
       = & \text{Tr}(U_k^\dagger U_l)=d_1\delta_{kl}.
  \end{array}
$$
Therefore, the $d_1^2$ states $\{U_j\otimes V|\alpha\rangle\}_{j=1}^{d_1^2}$ are pairwise orthogonal.
\qed

\begin{proposition}\label{unitary_extend} Let $\mathcal{L}_j $ be a set of distinguishable mutually orthogonal unitary matrices in $\text{\emph{U}}(d_j)$, $j=1,2,...,f$, and
$$
\widehat{\mathcal{L}}=\{U^{(1)}\otimes U^{(2)}\otimes\cdots \otimes U^{(f)} \ \big | \ U^{(j)}\in \mathcal{L}_j,~ j=1,2,...,f\}
$$
a set of unitary matrices of $\text{\emph{U}}(d_1d_2\cdots d_f),$ then, $\widehat{\mathcal{L}}$ are also distinguishable.
\end{proposition}

\noindent\emph{Proof:} Since $\mathcal{L}_j$ can be distinguished, there exists a unit vector $|\alpha_j\rangle\in \mathbb{C}^{d_j}$ such that $\{U|\alpha_j\rangle\ \ \big|\  U\in \mathcal{L}_j\}$ are pairwise orthogonal.
Set $|\alpha\rangle=|\alpha_1\rangle\otimes|\alpha_2\rangle\otimes\cdots\otimes|\alpha_f\rangle.$
Let $U=U^{(1)}\otimes U^{(2)}\otimes\cdots \otimes U^{(f)}$ and $V=V^{(1)}\otimes V^{(2)}\otimes\cdots \otimes V^{(f)}$ be two different elements in $\widehat{\mathcal{L}}$.
There must exist some $j_0$ such that $U^{(j_0)}$ and $V^{(j_0)}$ are different elements in $\mathcal{L}_{j_0}$,
$\langle \alpha_{j_0}|U^{(j_0)}{V^{(j_0)}}^\dagger|\alpha_{j_0}\rangle=0$, which implies that
$$\langle \alpha|UV^\dagger|\alpha\rangle=\prod_{j=1}^f\langle \alpha_{j}|U^{(j)}{V^{(j)}}^\dagger|\alpha_{j}\rangle=0.$$
Hence, $\{U|\alpha\rangle\ \big|\  U\in \widehat{\mathcal{L}}\}$ are indeed pairwise orthogonal, and the set $\widehat{\mathcal{L}}$ is distinguishable by definition.
\qed

From Proposition \ref{unitary_extend}, the remark in section \ref{sec}, and the Theorem \ref{main1}, we can easily deduce the following corollary.

\begin{corollary}\label{main2}
Let $\{|\psi_i^{(j)}\rangle\}_{i=0}^{d_j^2-1}$ be an orthonormal maximally entangled basis of $\mathcal{H}_{A_j}\otimes\mathcal{H}_{B_j}$ whose corresponding unitary matrices $\mathcal{B}_j=\{U_i^{(j)}\}_{i=0}^{d_j^2-1}$ are twist commutative for  $j=1,2,...,f$. Suppose that the states corresponding to $\mathcal{L}_j \subseteq \mathcal{B}_j$ can be distinguished by one-way LOCC from $A_j\rightarrow B_j$ for all $j$.
Then, the set corresponding to $\widehat{\mathcal{L}}=\{U^{(1)}\otimes U^{(2)}\otimes\cdots \otimes U^{(f)} \ \big | \ U^{(j)}\in \mathcal{L}_j, j=1,2,...,f\}$
can be also distinguished by one-way LOCC from $A\rightarrow B$ when considering the bipartite states as
$\mathcal{H}_A\bigotimes \mathcal{H}_B:=(\otimes_{j=1}^f\mathcal{H}_{A_j} )\bigotimes(\otimes_{j=1}^f\mathcal{H}_{B_j}).$
\end{corollary}

\vskip 8pt
\section{Application to qudit lattice states}\label{fiv}

In 2015, Tian \emph{et al.} successfully generalized Fan's result to $\mathbb{C}^{p^r}\otimes \mathbb{C}^{p^r}$ system \cite{Tian15}. In particular, they showed that any $l$ qudit lattice states in $\mathbb{C}^{p^r}\otimes \mathbb{C}^{p^r}$ are one-way locally distinguishable if $l(l-1)\leq 2p^r$. Based on these results and Corollary \ref{main2}, we can give a more general result for the arbitrary dimension $d$.

\begin{theorem}\label{lattice_distinguishability}
Let $d\geq 2$ be a positive integer with prime factorization $d=p_1^{r_1} p_2^{r_2}...p_f^{r_f} $, $p_1^{r_1}<p_2^{r_2}<\cdots <p_f^{r_f}.$ Then, any $l$ qudit lattice states in $\mathbb{C}^{d}\otimes \mathbb{C}^{d}$
are distinguishable by one-way LOCC provided that $l(l-1)\leq 2p_1^{r_1}.$
\end{theorem}

\noindent\emph{Proof:} Let $\mathcal{L}$ denote the corresponding unitary matrices of any given $l$ qudit lattice states. Every matrix in $\mathcal{L}$ can be written as  $f$ tensor products of the unitary matrices of dimension $p_i^{r_i}$ ($i=1,\cdots,f$), that is,
$$
U=U^{(1)}\otimes U^{(2)}\otimes \cdots \otimes U^{(f)},~ U^{(i)}\in \text{U}({p_i^{r_i}})
$$
for $i=1,2,...,f$. For each $i\in \{1,2,...,f\}$, we define
$$\mathcal{L}_i=\{U^{(i)}\ \ \big| \ U^{(1)}\otimes U^{(2)}\otimes \cdots \otimes U^{(f)}\in \mathcal{L}\}.$$
Let $l_i$ denote the number of elements in $\mathcal{L}_i$, $l_i:=|\mathcal{L}_i|\leq l$.
Then
$$
l_i(l_i-1)\leq l(l-1)\leq 2p_1^{r_1}\leq 2p_i^{r_i}.
$$
From Tian \emph{et al.}'s results in \cite{Tian15}, the states corresponding to $\mathcal{L}_i$ can be one-way LOCC distinguished.
Due to the necessity of Theorem \ref{main1}, the unitary matrices in $\mathcal{L}_i$ are distinguishable.
According to Corollary \ref{main2}, the unitary matrices in $\mathcal{L}\subseteq\widehat{\mathcal{L}}$ are also distinguishable.
From the sufficient part of Theorem \ref{main1}, the set of qudit lattice states corresponding to $\mathcal{L}$ are distinguishable by one-way LOCC.
\qed

\vskip 6pt
Let $l\geq 2$ be an integer. We say that $d$ satisfies $\mathcal{P}(l)$ if any $l$ unitary matrices corresponding to $l$ different qudit lattice states of dimension $d$
are distinguishable. Denote $P(l)=\{d\in\mathbb{N} \big | d\geq 2 \text{  and }d \text{ satisfied }  \mathcal{P}(l) \}$.
It is obvious that $P(l)\supseteq P(l+1)$.
As any two orthogonal bipartite states are distinguishable by one-way LOCC \cite{Walgate00}, it follows that $P(2)=\{2,3,4,...\}.$
It is interesting to determine $P(l)$ for $l\geq 3$.

\begin{theorem}\label{lattice_distinguishability_three}
Any three qudit lattice states in $\mathbb{C}^d\otimes \mathbb{C}^d$, $d\geq 3$ are distinguishable by one-way LOCC.
\end{theorem}

\noindent\emph{Proof:} Let $d=p_1^{r_1} p_2^{r_2}...p^{r_f}_f $ be the prime factorization of $d$ with $p_1^{r_1}<p_2^{r_2}<\cdots <p^{r_f}_f.$
If $p_1^{r_1}\geq 3(3-1)/2=3,$ we can draw the conclusion for such $d$ by Theorem \ref{lattice_distinguishability}.
Hence, we only need to consider the case $p_1^{r_1}=2$, that is, $d=2d'$ with $d'>1$ as $d\geq 3$(Moreover, 2 do not divide $d'$).
Suppose the unitary matrices corresponding to the three states are given by $$\mathcal{L}=\{U_i^{(1)}\otimes U_i^{(2)} | U_i^{(1)}\in \text{U}(2),U_i^{(2)}\in \text{U}(d'), i=1,2,3\}.$$
Denote $\mathcal{L}_1=\{U_i^{(1)}\ \big| i=1,2,3\}$ and $\mathcal{L}_2=\{U_i^{(2)}\ \big| i=1,2,3\}.$
\begin{enumerate}[(a)]
\item  $|\mathcal{L}_2|=1$. As $2< d'$, the condition of Proposition \ref{unitary_tele} is satisfied. Hence, $\mathcal{L}$ is distinguishable.
\item  $|\mathcal{L}_2|=2$. Without loss of generality, we can assume that $U_2^{(2)}=U_3^{(2)}.$  As $2,d'\in P(2)$, Then, there exist a unit vector $|\alpha_1\rangle \in \mathbb{C}^{2}$ and a unit vector $|\alpha_2\rangle \in \mathbb{C}^{d'}$ such that $\langle \alpha_1| {U^{(1)}_2}^\dagger {U^{(1)}_3}|\alpha_1\rangle=0$ and
      $$\langle \alpha_2| {U^{(2)}_1}^\dagger {U^{(2)}_2}|\alpha_2\rangle=\langle \alpha_2| {U^{(2)}_1}^\dagger {U^{(2)}_3}|\alpha_2\rangle=0.$$  Then for any $1\leq i\neq j\leq 3,$ we have
      $\langle\alpha|({U_i^{(1)}}^\dagger\otimes {U_i^{(2)}}^\dagger)( U_j^{(1)}\otimes U_j^{(2)})|\alpha\rangle=0$, i.e., the set $\mathcal{L}$ is also distinguishable.
  \item $|\mathcal{L}_2|=3$. In this case, $d'=p_2^{r_2}...p^{r_f}_f $ with  all $p_i^{r_i} \geq 3$ ($2\leq i\leq f$). By Theorem \ref{main1} and \ref{lattice_distinguishability}, the unitary matrices in $ \mathcal{L}_2$ are distinguishable. That is, there exists a unit vector $|\alpha_2\rangle \in \mathbb{C}^{d'}$ such that
  $$
  \langle \alpha_2| {U^{(2)}_i}^\dagger {U^{(2)}_j}|\alpha_2\rangle=\delta_{ij},~ \text{ for } 1\leq i,j\leq 3.
  $$
Choosing an arbitrary unit vector $|\alpha_1\rangle\in \mathbb{C}^{2}$ and setting $|\alpha\rangle=|\alpha_1\rangle\otimes |\alpha_2\rangle$,
we have the following relations
$$\langle\alpha|({U_i^{(1)}}^\dagger\otimes {U_i^{(2)}}^\dagger)( U_j^{(1)}\otimes U_j^{(2)})|\alpha\rangle=\delta_{ij},$$ which implies that $\mathcal{L}$ is distinguishable.
\end{enumerate}
Thus, the sufficient part of Theorem \ref{main1} implies the conclusion.
\qed

As any three Bell states cannot be locally distinguished, from Theorem \ref{lattice_distinguishability_three}, it follows that $P(3)=\{3,4,5,...\}.$
In the next theorem, we attempt to determine the $P(4)$.
\vskip 5pt

\begin{theorem}\label{lattice_distinguishability_4}
Any $4$ qudit lattice states in $\mathbb{C}^d\otimes \mathbb{C}^d$, $d\geq 91$ are distinguishable by one-way LOCC.
Furthermore,
$$\mathbb{N}\setminus\{1,2,3,4,5,6,10,12,15,18,20,30,50,60,90\}\subseteq P(4).$$
\end{theorem}

\noindent\emph{Proof:} We first present two claims, the proofs for which will be presented in Appendix A.
\begin{enumerate}[(a)]
  \item \textbf{Claim 1:} If $3\leq d'\leq d''$ and $d''\in P(4)$, then $d'd''\in P(4).$
  \item \textbf{Claim 2:} If $p$ is a prime number with $p\geq 7$, then $2p\in P(4)$.
\end{enumerate}
Now let $d=p_1^{r_1} p_2^{r_2}...p_f^{r_f} $ be the prime factorization of $d$ with $p_1^{r_1}<p_2^{r_2}<\cdots <p_f^{r_f}.$
If $p_1^{r_1}\geq 4(4-1)/2=6,$  by Theorems \ref{main1} and \ref{lattice_distinguishability}, it follows that $d\in P(4)$.
Hence, only those $d$ whose $p_1^{r_1}=2,3,4,5$ might lie outside $P(4)$.
If $f=1$, there are only four exceptional points, $d=2,3,4,5$. Therefore, we can assume $f\geq 2$.

If $p_1^{r_1}=3,4,5$, then $p_i^{r_i}$ cannot be greater than 6 for $ 2\leq i\leq f$. Otherwise, \textbf{Claim 1} and Proposition \ref{unitary_extend} would imply that $d\in P(4)$.
For example, $d=3\times 4\times 7\times 19^2$. Then $7,19^2\in P(4)$ imply that $7\times 19^2\in P(4)$ by Proposition \ref{unitary_extend}.
Hence, $4\times 7\times 19^2\in P(4)$ by \textbf{Claim 1}. So is $d$.
Therefore, this contributes four exceptional cases: $d=3\times 4, 3\times 5, 4\times5$, and $3\times 4\times 5.$

If $p_1^{r_1}=2$, then $p_f\leq5$. Otherwise, $2p_f\in P(4)$ by \textbf{Claim 2}. Using an argument similar to that above, we obtain $d\in P(4).$
Hence, $f=2$ or $3$. If $f=2$, then $d=2\times 3^i$ or $d=2\times 5^i$ with $i\leq 2$. Otherwise, we can decompose $d=6\times 3^{i-1}$ or $d=10\times 5^{i-1}$ and apply \textbf{Claim 1} to obtain $d\in P(4)$.
If $f=3$, then $d=2\times 3^i\times 5^j$ with $i\leq 2$ and $j=1.$
\qed

\section{Conclusion and Discussion}\label{last}
	
We have studied the problem of local distinguishability of MESs based on twist teleportation.
We have focused on the maximally entangled bases whose corresponding unitary matrices $\mathcal{S}=\{U_i\}_{i=1}^{d^2}$ are twist commutative, and we have shown that
the MESs perfectly corresponding to $\{U\}_{U\in \mathcal{L}}\subseteq \mathcal{S}$ are one-way LOCC distinguishable if and only if the set of unitary matrices $\{U\}_{U\in \mathcal{L}}$ are distinguishable.
By applying our results to the lattice qudit settings, we obtained a general version of Fan's and Tian \emph{et al.}'s results:
if $d=\prod_{i=1}^f p_i^{r_i}$ with $p_1^{r_1}<p_2^{r_2}<\cdots< p_f^{r_f}$, then any $l$ lattice states are one-way LOCC distinguishable if $l(l-1)/2\leq p_1^{r_1}.$
For given $l$, it is questionable whether any $l$ lattice states are LOCC distinguishable if the dimension $d$ is large enough.
However, using the main criterion that we obtained, we demonstrated that this is true for $l=3$ and $4$.
Some interesting questions remain, however; for example, how to determine whether a set of unitary matrices is distinguishable or not.
For $l\geq 5$, it is questionable whether there exist some $N(l)$ such that, if $d\geq N(l)$, then any $l$ qudit lattice states are distinguishable.
Furthermore, can one determine the set $P(l)$?
Our approach might merit further investigations regarding local distinguishability of bipartite or multipartite states.

\vspace{2.5ex}
\noindent{\bf Acknowledgments}\, \, This work was supported by the National Natural Science Foundation of
China (Grant Nos. 11675113, 11871295, and 11901084), Beijing Municipal
Commission of Education (Grant No. KM201810011009), Beijing Natural
Science Foundation (Grant No. Z190005), and the Research Startup Funds
of Dongguan University of Technology (Grant No. GC300501-103).

\vspace{2.5ex}
{
		$${\text{\textbf{APPENDIX A}}}$$
	}

\noindent\textbf{ Proof of Claim 1:} Suppose the unitary matrices corresponding to the four qudit lattice states are given by
$$
\mathcal{L}=\{U_i^{(1)}\otimes U_i^{(2)} | U_i^{(1)}\in \text{U}(d'),\,U_i^{(2)}\in \text{U}(d''),\, i=1,2,3,4\}.
$$
Denote
$$
\mathcal{L}_1=\{U_i^{(1)}\ \big| i=1,2,3,4\},\ \mathcal{L}_2=\{U_i^{(2)}\ \big| i=1,2,3,4\}.
$$
We separate the arguments into four cases according to the cardinality of $\mathcal{L}_2$.
\begin{enumerate}[(a)]
  \item $|\mathcal{L}_2|=1$. As $d'\leq d''$ the condition of Proposition 1 is fulfilled, hence $\mathcal{L}$ is distinguishable.
  \item $|\mathcal{L}_2|=2.$
 Without loss of generality, there are two subcases: i) $U^{(2)}_1=U^{(2)}_2=U^{(2)}_3\neq U^{(2)}_4$.
In this case, $U^{(1)}_1,U^{(1)}_2,U^{(1)}_3$ are pairwise different.
Since the three states are distinguished by one-way LOCC, there exists $|\alpha_1\rangle$ such that $\langle \alpha_1|{U^{(1)}_i}^\dagger{U^{(1)}_j}|\alpha_1\rangle =0$ for $1\leq i\neq j\leq 3$.
Clearly, there exists a unit vector $|\alpha_2\rangle$ such that $\langle \alpha_2|{U^{(2)}_i}^\dagger{U^{(2)}_4}|\alpha_2\rangle=0 $ for $i=1,2,3$.
Then $|\alpha\rangle=|\alpha_1\rangle\otimes|\alpha_2\rangle$ is just the vector we required.
ii) $U^{(2)}_1=U^{(2)}_2\neq U^{(2)}_3= U^{(2)}_4$. For this case, we need to find a vector $|\alpha_1\rangle$ such that
$\langle \alpha_1|{U^{(1)}_1}^\dagger{U^{(1)}_2}|\alpha_1\rangle =0$ and $\langle \alpha_1|{U^{(1)}_3}^\dagger{U^{(1)}_4}|\alpha_1\rangle =0.$
To show that the above conditions can be fulfilled, we define $\widetilde{\mathcal{L}}_1=\{I,{U_1^{(1)}}^\dagger{U_2^{(1)}},{U^{(1)}_3}^\dagger{U_4^{(1)}}\}.$
As $|\widetilde{\mathcal{L}}_1|\leq 3$, by Theorem 1 and Theorem 3, we can find a vector  $|\alpha_1\rangle$ such that
$\langle \alpha_1|U|\alpha_1\rangle=0$ for $U\in \Delta(\widetilde{\mathcal{L}}_1) $ (in this paper, we use the notation $\Delta(\mathcal{U}):=\{ U^\dagger V | U,V\in \mathcal{U}\}$ for any set $\mathcal{U}$ of unitary matrices of the same dimension and $\Delta(\mathcal{U})$ is called a pairwise difference set of $\mathcal{U}$  in   \cite{Tian15}).
Since both ${U_1^{(1)}}^\dagger{U_2^{(1)}}$ and ${U^{(1)}_3}^\dagger{U_4^{(1)}}$ are in $\Delta(\widetilde{\mathcal{L}}_1)$, we complete the proof.
   \item $|\mathcal{L}_2|=3$. We can assume $U^{(2)}_3=U^{(2)}_4$ without loss of generality. We then have $|\alpha_1\rangle$ and $|\alpha_2\rangle$
such that
$$\begin{array}{l}
  \langle\alpha_2| {U^{(2)}_i}^\dagger {U^{(2)}_j}|\alpha_2\rangle=0,~~ 1\leq i\neq j\leq 3, \\[2mm]
  \langle\alpha_1| {U^{(1)}_3}^\dagger {U^{(1)}_4}|\alpha_1\rangle=0.
\end{array}
$$
Then $|\alpha\rangle=|\alpha_1\rangle\otimes|\alpha_2\rangle$ is the vector we required.
 \item $|\mathcal{L}_2|=4$. As $\mathcal{L}_2$ is distinguishable, we can find  a vector $|\alpha_2\rangle$
such that
$$\begin{array}{c}
 \langle\alpha_2| {U^{(2)}_i}^\dagger {U^{(2)}_j}|\alpha_2\rangle=0,\ \ 1\leq i\neq j\leq 4. \\
\end{array}
$$
One can then choose an arbitrary unit vector $|\alpha_1\rangle\in \mathbb{C}^{d'}$ such that $|\alpha\rangle=|\alpha_1\rangle\otimes |\alpha_2\rangle$ is a vector we required. \qed
\end{enumerate}

\vskip 5pt

\noindent\textbf{ Proof of Claim 2:}
Let $d=2p$ and suppose that the unitary matrices corresponding to the four qudit lattice states are given by
$$\mathcal{L}=\{U_i^{(1)}\otimes U_i^{(2)} | U_i^{(1)}\in \text{U}(2),\,U_i^{(2)}\in \text{U}(p),\, i=1,2,3,4\}.$$ Denote
$$\mathcal{L}_1=\{U_i^{(1)}\ \big| i=1,2,3,4\},\ \mathcal{L}_2=\{U_i^{(2)}\ \big| i=1,2,3,4\}.$$
If $|\mathcal{L}_2|=1 \text{ or } 3 \text{ or } 4$, we can prove that $\mathcal{L}$ is distinguishable using a similar argument as that above.
We only need to consider the case $|\mathcal{L}_2|=2$.
In this case, up to a local unitary equivalence, we find that
{\small $$\Delta(\mathcal{L})\subseteq\{X_2\otimes I,  Y_2\otimes I, Z_2\otimes I, X_2\otimes X_p^l,Y_2\otimes X_p^l,Z_2\otimes X_p^l,I\otimes X_p^l\}$$}
for some $1\leq l\leq p-1$. Let $|\alpha\rangle=\sum_{i=0}^1\sum_{j=0}^{p-1} \alpha_{ij}|i\rangle|j\rangle.$ There exists some nontrivial solution $|\alpha\rangle$ such that
$\langle \alpha| U|\alpha\rangle=0$ for $U\in \Delta(\mathcal{L}).$ In fact, that
$$\begin{array}{l}
  \langle \alpha| X_2\otimes I|\alpha\rangle=0,  \langle \alpha| Y_2\otimes I|\alpha\rangle=0, \\[2mm]
  \langle \alpha| X_2\otimes X_p^l|\alpha\rangle=0,  \langle \alpha| Y_2\otimes X_p^l|\alpha\rangle=0, \\[2mm]
    \langle \alpha| Z_2\otimes X_p^l|\alpha\rangle=0,  \langle \alpha| I\otimes X_p^l|\alpha\rangle=0, \\[2mm]
    \langle \alpha| Z_2\otimes I_p|\alpha\rangle=0,
\end{array}
$$
is equivalent to that
$$\begin{array}{l}
   \sum_{j=0}^{p-1}\overline{\alpha_{0j}}\alpha_{1j}= \sum_{j=0}^{p-1}\overline{\alpha_{1j }}\alpha_{0j}=0, \\[2mm]
  \sum_{j=0}^{p-1}\overline{\alpha_{0j\oplus l}}\alpha_{1j}= \sum_{j=0}^{p-1}\overline{\alpha_{1j\oplus l}}\alpha_{0j}=0, \\[2mm]
   \sum_{j=0}^{p-1}\overline{\alpha_{0j\oplus l}}\alpha_{0j}= \sum_{j=0}^{p-1}\overline{\alpha_{1j\oplus l}}\alpha_{1j}=0, \\[2mm]
\sum_{j=0}^{p-1} |\alpha_{0j }|^2=\sum_{j=0}^{p-1} |\alpha_{1j }|^2,
\end{array}
$$
respectively. There exists some $1\leq j_0\leq p-1$ such that $j_0- l\neq0$ and $j_0+l\neq p$.
If we set
$$
\begin{array}{l}
 (\alpha_{00},\alpha_{01},...,\alpha_{0p-1})=(1,0,0,\cdots ,0 ,1,0,\cdots, 0, 0)\\
  (\alpha_{10},\alpha_{11},...,\alpha_{1p-1})=(1,0,0, \cdots ,0 ,-1,0,.\cdots, 0, 0)
 \end{array}$$
 (only the first and the $j_0$-th coordinates are nonzero),
then $|\alpha\rangle$ satisfies all the equations above. Hence $\mathcal{L}$ is distinguishable. \qed


\vspace{5ex}
	
	
\begin{thebibliography}{}
		
		\bibitem{Bennett99}
		 C. H. Bennett, D. P. DiVincenzo, C. A. Fuchs, T. Mor, E. Rains, P. W. Shor, J. A. Smolin, and W. K. Wootters,   \href{https://doi.org/10.1103/PhysRevA.59.1070}{Phys. Rev. A {\bf 59}, 1070 (1999)}.
		
		\bibitem{Walgate02}
J. Walgate and L. Hardy,   \href{https://doi.org/10.1103/PhysRevLett.89.147901}{Phys. Rev. Lett. {\bf 89}, 147901 (2002)}.
		
		\bibitem{DiVincenzo02}
			D. P. DiVincenzo, D.W.  Leung and B.M. Terhal,
		\href{https://ieeexplore.ieee.org/document/985948/}{IEEE Trans. Inf. Theory  \textbf{48}, 580 (2002)}.


\bibitem{Rahaman15}
 R. Rahaman,  M.G. Parker,   	\href{https://doi.org/10.1103/PhysRevA.91.022330}{Phys. Rev. A \textbf{91}, 022330 (2015).}

	
\bibitem{Ben99b} C. H. Bennett, D. P. DiVincenzo, T. Mor, P. W. Shor, J.
		A. Smolin,   B. M. Terhal,   \href{https://doi.org/10.1103/PhysRevLett.82.5385}{Phys. Rev. Lett. \textbf{82}, 5385
			(1999)}.
		
		
		\bibitem{Hor03} M. Horodecki, A. Sen(De), U. Sen, and K. Horodecki,
		\href{https://doi.org/10.1103/PhysRevLett.90.047902}{Phys. Rev. Lett. {\bf 90}, 047902 (2003)}.
		
		
	
		
		\bibitem{DiVincenzo03}  D. P. DiVincenzo, T. Mor, P. W. Shor, J. A. Smolin, and
		B. M. Terhal,   \href{https://doi.org/10.1007/s00220-003-0877-6}{Comm. Math. Phys. \textbf{238}, 379 (2003)}.


		\bibitem{Ran04}S. De Rinaldis,   \href{https://doi.org/10.1103/PhysRevA.70.022309}{Phys. Rev. A \textbf{70}, 022309 (2004)}.
		

	\bibitem{Feng09} Y. Feng and Y.-Y. Shi,   \href{https://doi.org/10.1109/TIT.2009.2018330}{IEEE Trans. Inf. Theory \textbf{55}, 2799 (2009)}.
		
		
		
		
		\bibitem{Yang13} Y.-H. Yang, F. Gao, G.-J. Tian, T.-Q. Cao, and Q.-Y.
		Wen,   \href{https://doi.org/10.1103/PhysRevA.88.024301}{Phys. Rev. A \textbf{88}, 024301 (2013)}.
	
		
		\bibitem{Zhang14P}  Z.-C. Zhang, F. Gao, G.-J. Tian, T.-Q. Cao and Q.-Y.
		Wen,  \href{https://doi.org/10.1103/PhysRevA.90.022313}{Phys. Rev. A \textbf{ 90}, 022313 (2014)}.
		
		\bibitem{Zhang15P} Z.-C. Zhang, F. Gao, S.-J. Qin, Y.-H. Yang, and Q.-Y. Wen,
		  \href{https://doi.org/10.1103/PhysRevA.92.012332}{Phys. Rev. A \textbf{92},
			012332 (2015)}.

\bibitem{Wang15} Y.-L. Wang, M.-S. Li, Z.-J. Zheng, and S.-M. Fei,   \href{https://doi.org/10.1103/PhysRevA.92.032313}{Phys. Rev. A \textbf{92},
			032313 (2015)}.	
	
		\bibitem{Zhang16P} Z.-C. Zhang, F. Gao, Y. Cao, S.-J. Qin, and Q.-Y. Wen,   \href{https://doi.org/10.1103/PhysRevA.93.012314}{ Phys. Rev. A \textbf{93}, 012314 (2016)}.
		
				
		\bibitem{Xu16m}
		G.-B. Xu, Q.-Y. Wen, S.-J. Qin, Y.-H. Yang, and F. Gao,   \href{https://doi.org/10.1103/PhysRevA.93.032341}{Phys. Rev. A \textbf{93}, 032341 (2016)}.
	
		
		\bibitem{Zhang17P}Z.-C. Zhang, K.-J. Zhang, F. Gao, Q.-Y. Wen, and C. H. Oh,
		 \href{https://doi.org/10.1103/PhysRevA.95.052344}{Phys. Rev. A \textbf{95}, 052344 (2017)}.
		
		
		\bibitem{Halder18} S. Halder,    \href{https://doi.org/10.1103/PhysRevA.98.022303}{Phys. Rev. A \textbf{98}, 022303 (2018)}.
		
		\bibitem{Walgate00}
		J. Walgate, A. J. Short, L. Hardy, and V. Vedral,
		\href{https://doi.org/10.1103/PhysRevLett.85.4972}{Phys. Rev. Lett. \textbf{85}, 4972 (2000)}.
		
			
		\bibitem{Ghosh01}
		S. Ghosh, G. Kar, A. Roy, A. Sen(De), and U. Sen,  \href{https://doi.org/10.1103/PhysRevLett.87.277902}{Phys. Rev. Lett. \textbf{87}, 277902 (2001)}.
			
		\bibitem{Ghosh04}
		S. Ghosh,  G. Kar, A. Roy  and D. Sarkar,
		\href{https://doi.org/10.1103/PhysRevA.70.022304}{ Phys. Rev. A \textbf{70}, 022304 (2004).}
		

		\bibitem{Duan12}
	N. Yu, R. Duan, and M. Ying,    \href{https://doi.org/10.1103/PhysRevLett.109.020506}{Phys. Rev. Lett. \textbf{109},
			020506 (2012)}.

	\bibitem{Cosentino13}
		A. Cosentino,   \href{https://doi.org/10.1103/PhysRevA.87.012321}{Phys. Rev. A \textbf{87}, 012321
			(2013)}.

	
		\bibitem{CosentinoR14}
		A. Cosentino  and   V. Russo,
		\href{http://vprusso.github.io/pdf/small_sets.pdf}{ Quantum Information \& Computation, \textbf{14}, 1098 (2014).}
		
	\bibitem{Li15}
M.-S. Li, Y.-L. Wang, S.-M. Fei and Z.-J. Zheng,   \href{https://doi.org/10.1103/PhysRevA.91.042318}{Phys.
			Rev. A \textbf{91}, 042318 (2015)}.
		
		\bibitem{Yu15}  S. X. Yu and C. H. Oh, \href{http://arxiv.org/abs/arXiv:1502.01274v1}{arXiv:1502.01274v1}.

\bibitem{Bandyopadhyay13}
S. Bandyopadhyay and M. Nathanson,   \href{https://journals.aps.org/pra/abstract/10.1103/PhysRevA.88.052313}{Phys. Rev. A \textbf{88}, 052313 (2013).}
		
		\bibitem{Yu14}
N. Yu, R. Duan, and M. Ying,
\href{https://ieeexplore.ieee.org/document/6747300/}{IEEE Trans. Inform. Theory \textbf{60}, 2069 (2014).}
		

\bibitem{Bandyopadhyay15}
S. Bandyopadhyay, A. Cosentino, N. Johnston, V. Russo, J. Watrous, and N. Yu,
\href{https://ieeexplore.ieee.org/document/7086052/}{IEEE Trans. Inform. Theory \textbf{61}, 3593 (2015).}
	
		
		\bibitem{Bandyopadhyay11}
	S. Bandyopadhyay, S. Ghosh, and G. Kar,   \href{https://doi.org/10.1088/1367-2630/13/12/123013}{New J. Phys.
			\textbf{13} 123013 (2011)}.
		
		\bibitem{Nathanson2013}
		M. Nathanson,
		\href{ https://journals.aps.org/pra/abstract/10.1103/PhysRevA.88.062316}{  Phys. Rev. A, \textbf{88}, 062316 (2013).}
		
		\bibitem{Zhang14}
		 Z.-C. Zhang,  Q.-Y. Wen,  F. Gao,  G.-J. Tian and  T.-Q. Cao,
		\href{https://link.springer.com/article/10.1007\%2Fs11128-013-0691-9}{  Quantum Inf. Proc. \textbf{13}, 795 (2014).}
		
		\bibitem{Zhang15}
		 Z.-C. Zhang,  K.-Q. Feng,  F.  Gao and  Q.-Y. Wen,
		 \href{https://journals.aps.org/pra/abstract/10.1103/PhysRevA.91.012329}{ Phys. Rev. A   \textbf{91}, 012329 (2015).}






\bibitem{Tian15_2} G.-J. Tian,   S.-X. Yu,  F. Gao,  Q.-Y. Wen and  C.H. Oh,		
	\href{https://journals.aps.org/pra/abstract/10.1103/PhysRevA.91.052314 }{Phys. Rev. A \textbf{91}, 052314 (2015).}
	

\bibitem{Yang18}Y.-H. Yang, J.-T. Yuan, C.-H.Wang,. S.-J. Geng, and H.-J. Zuo, \href{https://journals.aps.org/pra/abstract/10.1103/PhysRevA.98.042333}{\textcolor[rgb]{0.15,0.19,0.85}{Phys. Rev. A \textbf{98}, 042333 (2018)}}.
	
\bibitem{Wang19}	  Y.-L. Wang,  M.-S. Li,  Z.-X. Xiong, \href{https://journals.aps.org/pra/abstract/10.1103/PhysRevA.99.022307}{\textcolor[rgb]{0.15,0.19,0.85}{Phys. Rev. A \textbf{99}, 022307 (2019)}}.
	
\bibitem{Xiong19}	Z.-X. Xiong, M.-S. Li, Z.-J. Zheng, C.-J. Zhu, S.-M.  Fei, \href{https://journals.aps.org/pra/abstract/10.1103/PhysRevA.99.032346 }{\textcolor[rgb]{0.15,0.19,0.85}{Phys. Rev. A \textbf{99}, 032346  (2019)}}.

		\bibitem{Fan04}
	 H. Fan,   \href{https://doi.org/10.1103/PhysRevLett.92.177905}{Phys. Rev. Lett. \textbf{92}, 177905 (2004)}.
		
		\bibitem{Nathanson05}
M. Nathanson,  \href{https://doi.org/10.1063/1.1914731}{J. Math. Phys.   \textbf{46}, 062103 (2005)}.
		

		\bibitem{Tian15} G.-J. Tian,   S.-X. Yu,  F. Gao,  Q.-Y. Wen and  C.H. Oh,
		\href{https://journals.aps.org/pra/abstract/10.1103/PhysRevA.92.042320 }{Phys. Rev. A  \textbf{92}, 042320 (2015).}

		\bibitem{Tian16} G.-J. Tian,   S.-X. Yu,  F. Gao and  Q.-Y. Wen,
\href{https://journals.aps.org/pra/abstract/10.1103/PhysRevA.94.052315}{Phys. Rev. A \textbf{94}, 052315 (2016).}
		
		\bibitem{Singal15}
T. Singal, R. Rahaman, S. Ghosh, G. Kar,
	 	\href{https://journals.aps.org/pra/abstract/10.1103/PhysRevA.96.042314}{Phys. Rev. A  \textbf{96}, 042314 (2017).}
		
	\bibitem{Wang17}	
		 Y.-L. Wang,  M.-S. Li,  S.-M. Fei and  Z.-J. Zheng,   \href{https://link.springer.com/article/10.1007/s11128-017-1579-x}{Quantum Inf. Process.  \textbf{16}, 126 (2017).}

\bibitem{nil} M. A. Nielsen and I. L. Chuang, Quantum Computation and
Quantum Information(Cambridge University Press, Cambridge, U.K., 2004).


\bibitem{Werner01} R. F.  Werner,  \href{http://iopscience.iop.org/article/10.1088/0305-4470/34/35/332/meta }{J. Phys. A: Math. Gen. \textbf{34} (35): 7081 (2001).}





	
	\end{thebibliography}
\end{document}